\documentclass[a4paper,UKenglish,cleveref, autoref, thm-restate, nolineno]{socg-lipics-v2021}

\usepackage{mathtools}
\usepackage{amsmath}
\usepackage{algorithm}
\usepackage{amsfonts}
\usepackage{algpseudocode}

\pdfoutput=1 
\hideLIPIcs  

\nolinenumbers

\title{Exact and Efficient Sampling from Dynamic Discrete Distributions with Finite-Precision Weights}

\titlerunning{Exact Sampling from Dynamic Discrete Distributions} 

\author{Lilith Orion Hafner\footnote[1]{Equal Contribution}}{Independent Researcher, United States}{lilithhafner@gmail.com}{https://orcid.org/0009-0000-7858-5633}{}

\author{Adriano Meligrana\footnotemark[1]\footnote[2]{Corresponding author}}{Department of Computer, Control, and Management Engineering, Sapienza University of Rome, Italy}{adriano.meligrana@diag.uniroma1.it}{https://orcid.org/0009-0009-8317-528X}{}

\authorrunning{L. O. Hafner and A. Meligrana} 

\Copyright{Lilith Orion Hafner and Adriano Meligrana} 



\supplement{Source code}

\supplementdetails[
  linktext={WeightVectors.jl},
  subcategory={Algorithm Implementation}
]{Software}{https://github.com/LilithHafner/WeightVectors.jl}

\supplementdetails[
  linktext={DynamicDiscreteSamplersComparisons},
  subcategory={Benchmark Comparison}
]{Software}{https://github.com/ameligrana/DynamicDiscreteSamplersComparisons}

\EventEditors{John Q. Open and Joan R. Access}
\EventNoEds{2}
\EventLongTitle{42nd Conference on Very Important Topics (CVIT 2016)}
\EventShortTitle{CVIT 2016}
\EventAcronym{CVIT}
\EventYear{2016}
\EventDate{December 24--27, 2016}
\EventLocation{Little Whinging, United Kingdom}
\EventLogo{}
\SeriesVolume{42}
\ArticleNo{23}

\begin{document}

\maketitle

\begin{abstract}
Sampling from a dynamic discrete distribution means drawing an index with probability proportional to a mutable set of weights. Classical constant-time techniques such as the Alias Method \cite{walker1977} are well suited to static distributions, but become expensive in dynamic settings because updates require rebuilding auxiliary tables. Existing dynamic approaches, including Forest of Trees \cite{matias2003} and BUcket Sampling (BUS) \cite{zhang2023}, achieve reasonable practical performance but require infinite precision real arithmetic to be correct and produce meaningfully incorrect results when implemented on real hardware.

We present EBUS (Exact BUcket Sampling), a dynamic sampler for finite-precision weights that is exact by construction: every returned index has probability exactly proportional to its represented weight. Our guarantees are proved in a word RAM model with bounded exponent range. In that model, our method supports $O(1)$ worst-case expected sampling time, $O(1)$ amortized time to update a single weight, $O(n)$ space, and $O(n)$ construction. We also provide an implementation for IEEE 64-bit floating-point weights and show experimentally that it is competitive with, and often faster than, several implementations of previous inexact methods while avoiding their numerical failure modes.
\end{abstract}

\section{Introduction}
\label{sec:introduction}

Sampling from a discrete distribution is a classical problem: given indices $a_1, a_2, \dots, a_N$ with weights $w_1, w_2, \dots, w_N$, the goal is to sample $a_i$ with probability $w_i/\sum_{i=1}^{N}w_i$. Here we study the dynamic setting, where the indices and/or weights may change over time.

Efficient generation from dynamic discrete distributions is useful in, among other areas, database systems \cite{xie2021}, network analysis \cite{slepoy2008, allendorf2023}, bioinformatics \cite{colvin2010}, and complex systems science \cite{glielmo2025}.

We give a dynamic sampler that remains unbiased under arbitrary update policies and samples each index exactly according to its represented finite-precision weight. To the best of our knowledge, this is the first dynamic discrete sampler with exact finite-precision correctness, $O(1)$ worst-case expected sampling, and $O(1)$ amortized single-weight updates. Conceptually, EBUS is an exact finite-precision version of exponent bucketing: although this bucket structure was already analyzed in \cite{dambrosio2022, zhang2023}, we remove its dependence on infinite-precision real arithmetic while preserving efficiency.

Our results are stated in a word RAM model: the floating-point format is fixed in advance, so the number of exponent buckets is bounded by a constant. In that model, the algorithm has the following key properties:

\vspace{-5px}
\begin{itemize}
  \item \emph{Exactness}: the probability of selecting each element matches its weight exactly.
  \item \(O(1)\) \emph{worst-case expected} time per sample (with probabilistic termination).
  \item \(O(1)\) \emph{amortized} time to update any single weight.
  \item \(O(n)\) \emph{memory} to store the entire data structure.
  \item Low constant factors in practice: tens of cycles in typical cases, with pathological cases requiring only hundreds of cycles.
\end{itemize}

These features make it competitive with the most efficient previously known dynamic samplers \cite{zhang2023, allendorf2024, matias2003}, while avoiding any reliance on infinite-precision real arithmetic.

\section{Related Work}

For static distributions, the Alias Method \cite{walker1977} gives \(O(1)\) sampling after \(O(n)\) preprocessing. Another classical approach maintains cumulative sums in a binary tree \cite{wong1980}, yielding \(O(\log n)\) updates and draws. Static sampling has also been studied for space/query-time trade-offs on the word RAM \cite{bringmannlarsen2013}, for exact proportional and subset sampling in the static setting \cite{bringmannpanagiotou2017}, and for random-bit efficiency with the Fast Loaded Dice Roller (FLDR) \cite{saad2020}.

Dynamic sampling has a separate line of work. Rejection-based algorithms for changing distributions were studied in \cite{rajasekaranross1993} under assumptions based on a majorizing vector of per-item upper bounds. Later, \cite{hagerup1993} proved constant expected generation and update time for normalized weights in a randomized real RAM model, with analogous bounds for polynomially bounded integral weights. In a stronger arithmetic model with constant-time operations on arbitrary-size integers, \cite{matias2003} obtains \(O(\log^{*} n)\) expected update and generate times, together with an expected constant-time lookup-table refinement.

More recently, \cite{zhang2023} proposed BUS for dynamic weighted-set sampling. BUS groups elements into buckets by weight magnitude, samples a bucket according to its total weight, and then samples an element inside that bucket by rejection sampling. The scheme is formulated for positive real weights and analyzed without an explicit machine model such as real RAM or word RAM. It supports insertions and deletions in amortized \(O(1)\) time and draws a sample in \(O(\log n)\) time using linear space \cite{zhang2023}. On the algorithm-engineering side, \cite{allendorf2024} introduces the Proposal Array, with $O(n)$ construction, $O(1)$ expected sampling, and update time $O(\Delta w/\overline{w})$, where $\overline{w}$ is the mean weight.

These works pursue different goals. The static papers focus on space usage, preprocessing/query-time trade-offs, or entropy efficiency \cite{bringmannlarsen2013,bringmannpanagiotou2017,saad2020}. The dynamic papers study rejection-based schemes, use real-RAM or stronger arithmetic models, or provide practical bucket-based dynamic weighted-set sampling without strong guarantees \cite{rajasekaranross1993,hagerup1993,matias2003,zhang2023}. Our goal instead is exact proportional sampling from dynamically changing finite-precision machine weights: we work directly with their bit representations and prove exactness in a fixed word RAM model, without requiring infinite-precision arithmetic for correctness.

\section{Our Exact Algorithm}

All asymptotic bounds are with respect to the number $n$ of stored weights. We work in a word RAM model in which the floating-point format is fixed in advance, so the exponent range is bounded and arithmetic, bit operations, comparisons, and random generation on machine words take constant time. We assume that $n$ fits in one machine word. For simplicity, we assume a binary IEEE floating-point format that fits in one machine word of $b$ bits.

As in \cite{zhang2023}, we divide weights into exponent levels, first sample a level, and then sample an element within that level by rejection sampling. The difference is that we work directly with the finite-precision bit representation.

\subsection{Exact Inter-Level Sampling}

For a positive normal weight $w$, let $e(w)$ denote its biased exponent field and let $\sigma(w)$ denote its normalized integer significand, including the implicit leading bit and left shifted so that
\(
\sigma(w)\in[2^{b-1},2^b)
\).
This normalization is useful for the intra-level sampling described in the next section.

Exponent bucketing groups each weight $w$ by its exponent $e(w)$. Since subnormal numbers share the same zero exponent field, we first normalize them by shifting their significand, adding only a constant number of extra levels. Thus the data structure has a constant number $N$ of levels determined by the chosen IEEE format. For IEEE 64-bit floating point, there are 2046 normal-exponent levels and 52 normalized-subnormal levels, for a total of 2098.

For each level $L_\ell$, the data structure stores the exact significand sum $SS_\ell$ in a $2b$-bit integer, an approximate level weight $A_\ell$ in a $b$-bit integer for fast top-level scanning, and the list of indices currently in $L_\ell$ together with their normalized significands.

We use the $A_\ell$ values because directly sampling from $W_\ell = SS_\ell 2^{\ell}$ would require wide-integer arithmetic. Instead, we sample from the $A_\ell$ values and correct the rounding error with a low-probability rejection step. Once a level has been selected exactly, sampling within that level is also exact and constant expected time.

Each level $L_\ell$ contains exactly those weights whose normalized exponent is $\ell$. Since all weights in a level share the same exponent, its exact total weight is determined by the sum of its normalized significands:
\[
SS_\ell = \sum_{w\in L_\ell} \sigma(w)
\quad\text{and}\quad
W_\ell = SS_\ell 2^{\ell}.
\]

To avoid arbitrary-precision arithmetic in the common case, we also maintain a $b$-bit approximation $A_\ell$ of each level weight:
\[
A_\ell =
\begin{cases}
\lfloor W_\ell 2^G \rfloor + 1 & SS_\ell > 0,\\
0 & SS_\ell = 0.
\end{cases}
\]
where $G$ is a global shift adjusted during updates so that the total approximate mass $A=\sum_\ell A_\ell$ fits in one $b$-bit integer, hence in one machine word. We also define the total exact mass \(M=\sum_{\ell} W_\ell 2^G\). The shift $G$ is chosen to keep the corrective rejection step rare.

For a nonempty level $\ell$, write
\[
W_\ell 2^G = q_\ell + f_\ell,
\quad q_\ell = \lfloor W_\ell 2^G\rfloor,
\quad f_\ell \in [0,1).
\]
Then the maintained approximate bucket size is
\(
A_\ell = q_\ell + 1.
\)
Thus each nonempty level receives a bucket with $q_\ell$ ``safe'' points that are always accepted and one extra boundary point representing the discarded fractional mass $f_\ell$. Inter-level sampling therefore accepts immediately on the first $q_\ell$ points of the chosen bucket and invokes an exact correction only on the boundary point. That correction accepts with probability exactly $f_\ell$, so the accepted mass of the bucket is exactly $q_\ell+f_\ell = W_\ell 2^G$. More precisely, the sampler uses the following algorithm:

\begin{algorithm}[H]
\small
\caption{Exact inter-level sampling}\label{alg:inter-level}
\vspace{5px}
\vspace{5px}
\begin{algorithmic}[1]
\Procedure{Exact-Inter-Level-Sampling}{$(A_\ell, SS_\ell)_\ell, G$} \\
    \hspace*{\algorithmicindent} \textbf{Input:} approximate weights $A_\ell$, exact significand sums $SS_\ell$, global shift $G$\\
    \hspace*{\algorithmicindent} \textbf{Output:} a sampled level index $\ell$
    \State $x \gets \text{Unif}\bigl(\{1,\dots,\sum_\ell A_\ell\}\bigr)$
    \For{$\ell$ from the highest nonzero level down to 1}
        \If{$x < A_\ell$}
            \State \Return $\ell$
        \ElsIf{$x = A_\ell$}
            \State $s \gets G + \ell$
            \While{true}
                \State $s \gets s + b$
                \State $r \gets \text{Unif}\bigl(\{0,\dots,2^b-1\}\bigr)$
                \State $t \gets \lfloor SS_\ell 2^s \rfloor \bmod 2^b$
                \If{$r < t$}
                    \State \Return $\ell$
                \ElsIf{$r > t$ \text{or} $s \ge 0$}
                    \State restart the procedure 
                \EndIf
            \EndWhile
        \EndIf
        \State $x \gets x - A_\ell$
    \EndFor
\EndProcedure
\vspace{3px}
\end{algorithmic}
\end{algorithm}

All quantities used by Algorithm~\ref{alg:inter-level} are maintained incrementally during updates. Its hot loop uses only the $b$-bit approximate weights $A_\ell$ and enters a slow path only when the sampled integer lands on a bucket boundary. In that case it uses the exact $2b$-bit significand sum $SS_\ell$ and reveals the base-$2^b$ expansion of the fractional part lazily, $b$ bits at a time.

The boundary-refinement loop is a lazy comparison in base \(B=2^b\). When the sampled point lands on the extra boundary point of bucket \(\ell\), the algorithm must accept it with probability
\(
f_\ell=\{W_\ell 2^G\}\in[0,1).
\)
Since \(W_\ell=SS_\ell 2^\ell\) with \(SS_\ell\) an integer, \(W_\ell 2^G=SS_\ell 2^{\ell+G}\) has a terminating base-\(B\) fractional expansion, which the refinement loop reveals one digit at a time and compares with independent uniform base-\(B\) digits.

\begin{restatable}[Boundary-refinement subroutine]{lemma}{boundaryrefinement}
\label{lem:boundary-refinement}
Fix a nonempty level \(\ell\), let \(B=2^b\), and write
\[
W_\ell 2^G = q_\ell + f_\ell,
\quad
q_\ell = \lfloor W_\ell 2^G \rfloor,
\quad
f_\ell \in [0,1).
\]
Conditioned on reaching the boundary case \(x=A_\ell=q_\ell+1\) in
Algorithm~\ref{alg:inter-level}, the refinement loop accepts level \(\ell\) with probability exactly \(f_\ell\) and restarts the outer procedure with probability \(1-f_\ell\).
\end{restatable}

\begin{proof}
See Appendix \ref{sec:proofs}.
\end{proof}

\begin{restatable}[One-round acceptance probability for inter-level sampling]{lemma}{intersamplingprob}
\label{lem:inter-level-one-round}
Fix a shift $G$. In one outer iteration of Algorithm~\ref{alg:inter-level}, the probability of accepting level $\ell$ is exactly
\(
\frac{W_\ell 2^G}{A}.
\)
Consequently, the probability that the outer iteration accepts some level is
\(
P_{\mathrm{acc}} = \frac{M}{A}.
\)
\end{restatable}

\begin{proof}
See Appendix \ref{sec:proofs}.
\end{proof}

\begin{theorem}[Exactness of inter-level sampling]
\label{thm:inter-level-exactness}
Algorithm~\ref{alg:inter-level} returns level $\ell$ with probability exactly
\(
\frac{W_\ell}{\sum_j W_j}.
\)
\end{theorem}

\begin{proof}
By Lemma~\ref{lem:inter-level-one-round}, the probability that a given outer iteration accepts level $\ell$ is $W_\ell 2^G / A$, and the probability that it accepts some level is
\(
P_{\mathrm{acc}} = \frac{M}{A}.
\)
Therefore, conditioned on acceptance in that iteration, the probability of returning $\ell$ is
\[
\frac{W_\ell 2^G / A}{P_{\mathrm{acc}}}
=
\frac{W_\ell 2^G}{\sum_j W_j 2^G}
=
\frac{W_\ell}{\sum_j W_j}.
\]
Since successive outer iterations are independent and repeat until the first acceptance, the final output distribution is the same conditional distribution.
\end{proof}

The next corollary quantifies how the choice of $G$ affects the frequency of slow-path entries.

\begin{restatable}[Slow-path overhead]{corollary}{slowpathoverhead}
\label{cor:slow-path-overhead}
Let $N$ denote the number of possible levels, the expected number of times a full execution of Algorithm~\ref{alg:inter-level} enters the bucket-boundary refinement path is at most
\(
\frac{N}{M}.
\)
\end{restatable}

\begin{proof}See Appendix \ref{sec:proofs}.\end{proof}

\subsection{Exact Intra-Level Sampling}

Once a level has been chosen, we sample an element within that level by rejection sampling.

\begin{algorithm}[H]
\small
\caption{Exact intra-level sampling}\label{alg:intra-level}
\vspace{5px}
\vspace{5px}
\begin{algorithmic}[1]
\Procedure{Exact-Intra-Level-Sampling}{$(\sigma_j,j)_{j=1}^n$} \\
    \hspace*{\algorithmicindent} \textbf{Input:} the normalized significands of all elements in one level\\
    \hspace*{\algorithmicindent} \textbf{Output:} an index $j \in \{1,\dots,n\}$
    \State $B \gets 2^{\lceil \log_2 n \rceil}$
    \While{true}
        \State $j \gets \text{Unif}(\{1,\dots,B\})$
        \State $x \gets \text{Unif}(\{0,\dots,2^b-1\})$
        \If{$j \le n$ \textbf{and} $x < \sigma_j$}
            \State \Return $j$
        \EndIf
    \EndWhile
\EndProcedure
\vspace{3px}
\end{algorithmic}
\end{algorithm}

Because all elements in a fixed level share the same exponent, exactness reduces to showing that Algorithm~\ref{alg:intra-level} samples indices in proportion to their normalized significands. The runtime bound follows because all normalized significands lie in the upper half of the $b$-bit range.

\begin{theorem}[Exactness of intra-level sampling]
\label{thm:intra-level-exactness}
Let a fixed level contain elements with normalized significands $\sigma_1,\dots,\sigma_n$. Algorithm~\ref{alg:intra-level} returns index $j$ with probability exactly
\(
\frac{\sigma_j}{\sum_{k=1}^n \sigma_k}.
\)
Equivalently, since all weights in the level share the same exponent, it returns each element with probability proportional to its weight within that level.
\end{theorem}

\begin{proof}
Let $B=2^{\lceil\log_2 n\rceil}$. In one iteration of the loop, index $j$ is proposed with probability $1/B$ and accepted with probability $\sigma_j/2^b$. Hence the probability that one iteration accepts index $j$ is
\(
\frac{1}{B}\cdot\frac{\sigma_j}{2^b}.
\)
The probability that one iteration accepts some index is therefore
\[
\sum_{k=1}^n \frac{1}{B}\cdot\frac{\sigma_k}{2^b}
=
\frac{1}{B2^b} \sum_{k=1}^n \sigma_k.
\]
Conditioned on acceptance in that iteration, the returned index has distribution
\[
\frac{\frac{1}{B}\cdot\frac{\sigma_j}{2^b}}{\frac{1}{B2^b} \sum_{k=1}^n \sigma_k}
=
\frac{\sigma_j}{\sum_{k=1}^n \sigma_k}.
\]
As in the inter-level case, repeating independent iterations until the first acceptance preserves this conditional distribution.
\end{proof}

\begin{restatable}[Expected number of intra-level proposals]{corollary}{intraproposals}
\label{cor:intra-level-runtime}
The expected number of iterations of Algorithm~\ref{alg:intra-level} is at most $4$.
\end{restatable}

\begin{proof}
See Appendix \ref{sec:proofs}.
\end{proof}

\subsection{Global shift adjustments}

For a global shift \(G\), each nonempty level \(\ell\) stores
\(
A_\ell(G)=\lfloor W_\ell 2^G\rfloor+1
\)
with \(W_\ell = SS_\ell 2^\ell\), while empty levels store \(A_\ell(G)=0\). Let
\(
A(G)=\sum_{\ell} A_\ell(G),\:
M(G)=\sum_{\ell} W_\ell 2^G,\:
z=\bigl|\{j:w_j>0\}\bigr|,
\)
and assume \(z<2^b\). Since
\[
A(G)-N \le M(G) \le A(G),
\]
it suffices to control \(A(G)\) for the slow-path overhead of Algorithm~\ref{alg:inter-level}. The shift affects efficiency, not exactness, and is chosen so that the total approximate mass both fits in one $b$-bit word and remains large enough for efficient sampling. This leads to the following definition:

\begin{definition}[Safe, good, and strongly good shifts]
\label{def:safe-good-shifts}
Fix the machine-word size \(b\) and target exponents \(K\le L\).
A shift \(G\) is called \emph{safe} if
\(
A(G)<2^b,
\)
\emph{good} if
\(
A(G)\ge 2^K,
\)
and \emph{strongly good} if
\(
A(G)\ge 2^L.
\)
\end{definition}

\begin{corollary}[Slow-path overhead for good and strongly good shifts]
\label{cor:good-shift-overhead}
If \(G\) is good, then the expected number of times a completed execution of
Algorithm~\ref{alg:inter-level} enters the refinement path is at most
\(
\frac{N}{2^K-N}.
\)
If \(G\) is strongly good, then this bound improves to
\(
\frac{N}{2^L-N}.
\)
\end{corollary}

\begin{proof}
If \(G\) is good, then \(A(G)\ge 2^K\), hence \(M(G)\ge A(G)-N\ge 2^K-N\). Corollary~\ref{cor:slow-path-overhead} therefore gives
\(
\mathbb{E}[S]\le \frac{N}{M(G)}\le \frac{N}{2^K-N}.
\)
The strongly good case is identical, using \(A(G)\ge 2^L\).
\end{proof}

The global-adjustment policy uses a lazy increase that estimates the total exact mass relative to the largest nonempty level. If \(\lambda\) is that level, define
\(
Y_\lambda=\sum_{r\ge 0} SS_{\lambda-r}2^{-r+1-b},
\)
then
\(
M(G)=2^{G+\lambda+b-1}Y_\lambda.
\)
So choosing \(G\) reduces to estimating the order of magnitude of \(Y_\lambda\). The next lemma shows that the coarse estimator is accurate up to an additive error depending only on \(b\).

\begin{restatable}[Quality of the coarse estimator]{lemma}{coarseestimator}
\label{lem:coarse-estimator}
Let \(\lambda\) be the largest nonempty level, and for \(r=0,1,\dots,b\), define
\[
H_r=\left\lfloor \frac{SS_{\lambda-r}}{2^b}\right\rfloor,
\quad
E=\sum_{r=0}^{b}\left\lfloor H_r2^{-r+1}\right\rfloor.
\]
Then
\(
E \le Y_\lambda < E+b+7.
\)
Moreover, if the sampler is nonempty, then
\(
Y_\lambda \ge 1.
\)
\end{restatable}
\begin{proof}
See Appendix \ref{sec:proofs}.
\end{proof}

The next lemma provides the basic estimate used to analyze the downward shift steps of the policy.

\begin{restatable}[Downward-shift bounds]{lemma}{downwardbounds}
\label{lem:downshift-bounds}
Let \(d\ge1\), suppose \(G'=G-d\), and let \(S\) be any set of nonempty levels. Then
\[
\sum_{r\in S} A_r(G')
\le
|S|+\left\lfloor \frac{\sum_{r\in S}A_r(G)-|S|}{2}\right\rfloor.
\]
If \(S\) is the set of all nonempty levels, then
\(
A(G')
\le
N+\left\lfloor \frac{A(G)}{2^d}\right\rfloor.
\)
\end{restatable}
\begin{proof}
See Appendix \ref{sec:proofs}.
\end{proof}

We now state the policy for changing \(G\). Any resulting repair touches only the top \(2b\) potentially nontrivial levels; Appendix \ref{sec:restoration-levels} explains how to restore them efficiently.

\begin{theorem}[Global-shift policy]
\label{thm:global-shift-adjustments}
Fix the machine-word size \(b\), the number of levels \(N\), and parameters
\(
K,\; \Delta,\; R,\; L_0,\; L \in \mathbb{N^+}
\)
satisfying
\[
2 \le \Delta \le R - K < b-K-2,
\quad
\log_2{N}< K \le L_0\le L < b - \log_2(b+10).
\]
The sampler updates the global shift \(G\) by the following rules.

\begin{enumerate}
    \item \textbf{Initialization.}
    If the sampler is empty and the first nonzero weight is inserted into level \(\lambda\), set
    \(
    G=L_0+1-b-\lambda.
    \)

    \item \textbf{Decrease when level overflow.}
    If an update would change level \(\ell\) to a new positive exact sum \(SS_\ell\) and
    \(
    \lfloor W_\ell 2^G\rfloor+1 \ge 2^b,
    \)
    set
    \(
    G_{\downarrow}=R-\lfloor \log_2 SS_\ell\rfloor-\ell.
    \)

    \item \textbf{Decrease when global overflow.}
    Otherwise, if the total approximate mass would overflow at the current shift, set
    \(
    G_{\downarrow} \leftarrow G-\Delta.
    \)

    \item \textbf{No eager increase after deletions.}
    If an update decreases a level sum or empties a level, leave \(G\) unchanged.

    \item \textbf{Lazy increase before sampling.}
    Before Algorithm~\ref{alg:inter-level}, if
    \(
    A(G)<2^K,
    \)
    let \(\lambda\) be the largest nonempty level, let \(E\) be as in Lemma~\ref{lem:coarse-estimator}, and set
    \(
    G_{\uparrow}=L+1-b-t-\lambda
    \)
    where \(
    t=\lfloor \log_2(\max\{E,1\})\rfloor
    \).
\end{enumerate}
\vspace{-3px}
Then:
\vspace{-3px}
\begin{enumerate}
    \item Initialization produces a safe and good shift.

    \item Every eager decrease, i.e. step~2 or step~3 during an insertion, produces a safe and good shift.

    \item Every deletion or downward update preserves safety, though goodness and strong goodness may be lost. Whenever a later sampling call finds \(A(G)<2^K\), step~5 restores a safe and strongly good shift.
\end{enumerate}

Consequently, after restoring the stored summaries to the exact values \(A_\ell(G)\) for the new shift, all previous exactness statements remain valid.
\end{theorem}

\begin{proof}
We verify the three claims.

\emph{Initialization.}
If the first nonzero weight is inserted into level \(\lambda\), then
\(SS_\lambda\in[2^{b-1},2^b)\). With
\(
G=L_0+1-b-\lambda,
\)
we have
\(
M(G)=W_\lambda 2^G=SS_\lambda 2^{L_0+1-b}\in[2^{L_0},2^{L_0+1}).
\)
Since only one level is nonempty,
\(
A(G)=\lfloor M(G)\rfloor+1<2^{L_0+1}+1\le 2^b,
\)
so the shift is safe, and
\(
A(G)=\lfloor M(G)\rfloor+1\ge M(G)\ge 2^{L_0}\ge 2^K.
\)
Thus it is good.

\emph{Eager decreases after insertions.}
Let \(G\) be the shift before the insertion, and let \(\ell\) be the touched level.
The pre-update state is safe, so \(A(G)<2^b\). There are two cases.

\emph{Case 1: the touched bucket would overflow at \(G\).}
Assume
\(
\lfloor W_\ell 2^{G}\rfloor+1\ge 2^b.
\)
Then step~2 sets
\(
G_{\downarrow}=R-\lfloor\log_2 SS_\ell\rfloor-\ell.
\)
By construction,
\(
W_\ell 2^{G_{\downarrow}}
=
SS_\ell\,2^{R-\lfloor\log_2 SS_\ell\rfloor}
\in [2^R,2^{R+1}),
\)
so
\(
M(G_{\downarrow})\ge W_\ell 2^{G_{\downarrow}}\ge 2^R\ge 2^K.
\)
Hence \(G_{\downarrow}\) is good because \(A(G_{\downarrow})\ge M(G_{\downarrow})\). For safety, since \(R<b-2\),
\(
W_\ell 2^{G_{\downarrow}}<2^{R+1}\le 2^{b-2}<2^b-1.
\)
The overflow assumption gives
\(
W_\ell 2^{G}\ge 2^b-1.
\)
Thus \(G_{\downarrow}<G\). Write
\(
d=G-G_{\downarrow}\ge 1,
\)
and
\(
S=\{r\neq \ell : SS_r>0\},
\)
and apply Lemma~\ref{lem:downshift-bounds} to the untouched nonempty levels:
\[
\sum_{r\neq \ell}A_r(G_{\downarrow})
\le
|S|+\left\lfloor \frac{\sum_{r\neq \ell}A_r(G)-|S|}{2}\right\rfloor.
\]
Since the pre-update state was safe,
\(
\sum_{r\neq \ell}A_r(G)<2^b,
\)
and \(|S|<N\),
\(
\sum_{r\neq \ell}A_r(G_{\downarrow})<N+2^{b-1}.
\)
Also,
\(
A_\ell(G_{\downarrow})=\lfloor W_\ell 2^{G_{\downarrow}}\rfloor+1<2^{R+1}+1.
\)
Therefore
\[
A(G_{\downarrow})
=
A_\ell(G_{\downarrow})+\sum_{r\neq \ell}A_r(G_{\downarrow})
<
N+2^{b-1}+2^{R+1}+1.
\]
Now the assumptions imply
\(
N<2^K\le 2^{b-2},
\)
while \(R<b-2\) implies
\(
2^{R+1}\le 2^{b-2}.
\)
Hence
\[
A(G_{\downarrow})
<
N+2^{b-1}+2^{R+1}+1
\le
(2^{b-2}-1)+2^{b-1}+2^{b-2}+1
=
2^b.
\]
So \(G_{\downarrow}\) is safe.

\emph{Case 2: no individual bucket would overflow, but the total does.}
Assume that
\(
\lfloor W_\ell 2^{G}\rfloor+1<2^b
\)
but
\(
A(G)\ge 2^b
\)
after the insertion. Then step~2 is skipped and step~3 sets
\(
G_{\downarrow}=G-\Delta.
\)
For safety, the insertion changes only level \(\ell\), so, since the pre-update state was safe,
\(
\sum_{r\neq \ell}A_r(G)<2^b,
\)
and by assumption the touched bucket is also \(<2^b\). Hence
\(
A(G)<2^{b+1}.
\)
Applying Lemma \ref{lem:downshift-bounds} with \(d=\Delta\) yields
\[
A(G_{\downarrow})\le N+\left\lfloor\frac{A(G)}{2^\Delta}\right\rfloor
<
N+2^{b+1-\Delta}
< 2^{b-2}+2^{b-1} <
2^b,
\]
so \(G_{\downarrow}\) is safe. For goodness, let \(m\) be the number of nonempty levels after the insertion, with \(m\le N\). For every nonempty level \(r\),
\(
W_r2^{G}\ge A_r(G)-1,
\)
because \(A_r(G)=\lfloor W_r2^{G}\rfloor+1\). Summing over all nonempty levels gives
\[
M(G)\ge \sum_{r:\,SS_r>0}(A_r(G)-1)=A(G)-m\ge 2^b-N.
\]
Therefore
\[
M(G_{\downarrow})=2^{-\Delta}M(G)\ge \frac{2^b-N}{2^\Delta}\ge 2^K,
\]
where the last inequality follows since
\(
2^{K+\Delta}+N\le 2^{K+\Delta}+2^K\le 2^b
\)
by hypothesis. Thus \(G_{\downarrow}\) is good.

\emph{Deletions and downward updates.}
If an update decreases a level sum while leaving \(G\) unchanged, then every
\(A_\ell(G)\) weakly decreases, so safety is preserved. However, \(A(G)\) may fall below \(2^K\) or \(2^L\), so goodness and strong goodness may be lost.

\emph{Lazy increase before sampling.}
Let \(\lambda\) be the largest nonempty level, and define \(Y_\lambda\), \(E\), and \(t=\lfloor \log_2(\max\{E,1\})\rfloor\)
as in the statement. By Lemma~\ref{lem:coarse-estimator},
\[
2^t\le Y_\lambda < E+b+7 \le \max\{E,1\}+b+7 < 2^{t+1}+b+7.
\]
With
\(
G_{\uparrow}=L+1-b-t-\lambda,
\)
we have
\[
M(G_{\uparrow})=2^{G_{\uparrow}+\lambda+b-1}Y_\lambda=2^{L-t}Y_\lambda \ge 2^{L-t}2^t=2^L,
\]
so the new shift is strongly good. Also,
\[
M(G_{\uparrow})
<
2^{L-t}(2^{t+1}+b+7)
=
2^{L+1}+(b+7)2^{L-t}
\le (b+9)2^L.
\]
Hence
\[
A(G_{\uparrow})\le M(G_{\uparrow})+N<(b+9)2^L+N<(b+10)2^L<2^b.
\]
So it is safe as well. This proves the claims.
\end{proof}

Our implementation instantiates the policy with
\[
b=64,
\quad
N=2098,
\quad
L_0=39,
\quad
R=48,
\quad
K=32,
\quad
L=46,
\quad
\Delta=16.
\]
These choices make the initial state good, preserve goodness after every eager decrease, and ensure that after a lazy increase
\(
2^{46} \le A(G) < (64+10)2^{46} < 2^{53}
\).
Hence Corollary~\ref{cor:good-shift-overhead} yields
\(
\frac{2098}{2^{32}-2098} < 2^{-20}
\)
for good shifts and
\(
\frac{2098}{2^{46}-2098} < 2^{-34}
\)
for strongly good shifts.

We next show why a shift change never requires touching all \(N\) levels: sufficiently small levels are forced to be unit buckets and do not need any update.

Assume the sampler is nonempty and let \(\lambda\) denote the largest nonempty level. For every shift \(G\), define
\(
\theta(G,z) = -b-G-\lfloor \log_2(z) \rfloor
\).
The threshold \(\theta(G,z)\) marks the first level that is not forced to be a unit bucket as proved in the next lemma.

\begin{lemma}[Levels below a threshold are unit buckets]
\label{lem:unit-bucket-threshold}
Assume the sampler is nonempty. If
\(
\ell < \theta(G,z) = -b-G-\lfloor \log_2(z)\rfloor,
\)
then
\[
A_\ell(G)=
\begin{cases}
0, & SS_\ell=0,\\
1, & SS_\ell>0.
\end{cases}
\]
\end{lemma}

\begin{proof}
Each stored normalized significand is strictly less than \(2^b\), so summing at most \(z\) of them gives
\(
SS_\ell < z\,2^b.
\)
If \(\ell< -b-G-\lfloor \log_2(z) \rfloor\), then
\(
\ell+G \le -b-1-\lfloor \log_2(z) \rfloor,
\)
hence
\[
W_\ell 2^G
=
SS_\ell 2^{\ell+G}
<
z\,2^b\,2^{-b-1-\lfloor \log_2(z) \rfloor}
\le 1.
\]
Thus \(\lfloor W_\ell 2^G\rfloor=0\), and the stated formula follows.
\end{proof}

\begin{corollary}[Low levels are unaffected by a safe shift change]
\label{cor:low-levels-unchanged}
Suppose the policy changes the shift from \(G\) to \(G'\). Then every level satisfying
\(
\ell < \min\{\theta(G,z),\theta(G',z)\}
\)
obeys
\(
A_\ell(G)=A_\ell(G')\in\{0,1\}
\).
Hence these levels require no summary update.
\end{corollary}

\begin{proof}
Apply Lemma~\ref{lem:unit-bucket-threshold} at both shifts. Every level below the smaller threshold is a unit bucket before and after the change.
\end{proof}

\begin{theorem}[Max \(2b\) approximate levels' updates per shift adjustment]
\label{thm:max-updated-levels}
Suppose the policy changes the shift from \(G\) to \(G'\). Then at most \(2b\) level summaries can require updating.
\end{theorem}

\begin{proof}
By Corollary~\ref{cor:low-levels-unchanged}, only levels
\(
\ell \ge \min\{\theta(G,z),\theta(G',z)\}
\)
can change. Let \(H=\max\{G,G'\}\). Then
\(
\min\{\theta(G,z),\theta(G',z)\}=\theta(H,z).
\)
Because the shift change occurs according to the policy, both the old and new shift are safe by Theorem~\ref{thm:global-shift-adjustments}. In particular, \(H\) is safe, so \(A(H)<2^b\). Because level \(\lambda\) is nonempty, it contains at least one normalized significand, hence
\(
SS_\lambda \ge 2^{b-1}.
\)
Therefore
\[
SS_\lambda 2^{\lambda+H}< A_\lambda(H) = \lfloor SS_\lambda 2^{\lambda+H}\rfloor+1\le A(H)<2^b.
\]
Using \(SS_\lambda\ge 2^{b-1}\), we obtain
\(
2^{b-1}2^{\lambda+H}< 2^b,
\)
hence
\(
\lambda+H<1.
\)
Since \(\lambda+H\) is an integer, \(\lambda+H\le 0\). Also, because \(z<2^b\), we have
\(
\lfloor \log_2 z\rfloor \le b-1.
\)
Therefore
\[
\theta(H,z)
=
-b-H-\lfloor \log_2 z\rfloor
\ge
-b-H-(b-1)
\ge
\lambda-2b+1.
\]
So the number of levels in the interval \([\theta(H,z),\lambda]\) is at most
\(
\lambda-\theta(H,z)+1 \le 2b.
\)
\end{proof}

With this knowledge, we can also provide a bound on the expected number of levels to be linearly scanned in one outer sampling iteration which is almost always better than the maximum number of levels $N$:

\begin{restatable}[Expected scan length of one outer iteration]{corollary}{scanlength}
\label{cor:one-outer-iteration-scan}
In one outer iteration of Algorithm~\ref{alg:inter-level}, the expected number of scanned levels is at most
\(
2b+\frac{N(N+1)}{2^{K+1}}.
\)
\end{restatable}

\begin{proof}
See Appendix \ref{sec:proofs}.
\end{proof}
\subsection{Updates and Amortization}

In this section, we analyze how a weight update affects the local level data structures, omitting the global shift guarantees from the previous section.

An update to one stored weight is decomposed into at most two operations: deleting the old nonzero weight, if any, and inserting the new nonzero weight, if any. We describe the mechanism using a per-level array representation.

For each normalized level \(\ell\), we maintain a dynamic array \(V_\ell\).  Each live entry of \(V_\ell\) is a pair \((\sigma_j,j)\), where \(j\) is the logical index of a stored weight in that level and \(\sigma_j\) is its normalized significand.  A separate locator table stores, for each live index \(j\), the pair \((\ell,p)\) such that \((\sigma_j,j)\) is currently the \(p\)th entry of \(V_\ell\).

Suppose index \(j\) currently lies at position \(p\) of level \(\ell\). To delete it, we swap the last entry of \(V_\ell\) into position \(p\), update the locator of the moved index if necessary, remove the last entry, and subtract \(\sigma_j\) from \(SS_\ell\). If the new exact sum is zero, we set \(A_\ell=0\); otherwise we recompute \(A_\ell=\lfloor W_\ell 2^G\rfloor+1\) from the updated exact sum and the current shift \(G\). To insert a new nonzero weight of normalized significand \(\sigma_j\) into level \(\ell\), we append \((\sigma_j,j)\) to \(V_\ell\), record its locator, add \(\sigma_j\) to \(SS_\ell\), and update the bucket size \(A_\ell=\lfloor W_\ell 2^G\rfloor+1\). If a weight changes from one nonzero value to another, we perform one deletion followed by one insertion.

The only non-worst-case aspect of the update procedure is resizing the per-level dynamic arrays \(V_\ell\). We assume standard geometric resizing: when a vector is full, its capacity doubles; optionally, when its occupancy falls below a fixed fraction \(t<1/2\), it may shrink to capacity \(O(|V_\ell|)\). By the standard analysis of dynamic arrays, over any sequence of \(U\) logical updates, the total number of element moves caused by resizing is \(O(U)\). Thus resizing contributes only \(O(1)\) amortized time per update. The total number of stored pairs across all levels is exactly the number of nonzero weights, and geometric resizing keeps the total reserved capacity within a constant factor of that quantity, so the update structure uses \(O(n)\) space. For the theory, this per-level array-of-vectors representation is sufficient. Our implementation instead packs all per-element records into one manually managed global array together with per-level ranges; this changes only constant factors and memory locality.

\subsection{Correctness and Performance Guarantees}

We conclude by summarizing the main sampler's correctness and performance guarantees.

\begin{theorem}[Correctness and performance guarantees of EBUS]
\label{thm:main-guarantee}
In the word RAM model, EBUS samples exactly from the represented finite-precision weights. It supports \(O(1)\) worst-case expected time per sample, \(O(1)\) amortized time per weight update, \(O(n)\) space, and \(O(n)\) construction from an input vector of length \(n\).
\end{theorem}

\begin{proof}
Exactness follows from Theorems~\ref{thm:inter-level-exactness},
\ref{thm:intra-level-exactness}, and \ref{thm:global-shift-adjustments}. For sampling, if the current shift \(G\) is good, Corollary~\ref{cor:good-shift-overhead} bounds the expected number of refinement-path entries in a completed execution of Algorithm~\ref{alg:inter-level} by \(N/(2^K-N)=O(1)\). Every restart occurs only after such an entry, so the expected number of outer iterations is also \(O(1)\). Each iteration scans at most \(N\) levels, and \(N\) is constant in the word RAM model; hence the inter-level sampler is \(O(1)\) expected time for good shifts. If \(G\) is not good, then \(A(G)<2^K\), so by Theorem~\ref{thm:global-shift-adjustments} a lazy increase is performed immediately before sampling and restores a strongly good, hence good, shift. Therefore the inter-level sampler is \(O(1)\) worst-case expected time. The refinement routine has constant expected cost, and the intra-level sampler has constant expected cost by Corollary~\ref{cor:intra-level-runtime}; thus one sample is produced in \(O(1)\) worst-case expected time.

For updates, one logical update consists of at most one deletion and one insertion, so at most two levels are touched directly. If a shift adjustment occurs, Theorem~\ref{thm:max-updated-levels} shows that at most \(2b\) level summaries need repair, each in \(O(1)\) time; since \(b\) and the number of levels \(N\) are fixed, all non-resizing work is worst-case \(O(1)\). Geometric resizing contributes \(O(1)\) amortized time, so updates are \(O(1)\) amortized.

The structure stores \(O(1)\) metadata per level and dynamic arrays of total capacity \(O(n)\), giving \(O(n)\) space. Construction scans the input once and performs at most one insertion per entry, so it is \(O(n)\).
\end{proof}

\section{Experiments}

To evaluate numerical accuracy and performance, we benchmarked EBUS against previous dynamic samplers\footnote{All benchmarks were executed on an x86\_64 Linux system with an AMD Ryzen 5-5600H CPU and 16~GB of RAM.}. We selected representative implementations of the main competing approaches:

\vspace{-5px}
\begin{itemize}

    \item our Julia implementation of EBUS;
    
    \item the C++ BUS implementation of \cite{zhang2023}, which could be considered the inexact version of our method;

    \item the Rust implementation of \cite{matias2003} used in \cite{allendorf2023} for nonlinear preferential attachment. We call it FT (Forest of Trees), following the terminology of \cite{matias2003};
    
    \item the C++ implementation of the Dynamic Proposal Array* (DPA*) method of \cite{allendorf2024}. An algorithm which was shown to be the most performant when compared to other methods by its author.
\end{itemize}

\subsection{Numerical Comparison}

To compare numerical accuracy, we fixed a set of $n$ items and gradually decayed their weights over $t$ iterations. At each decay step, we drew many samples and compared the empirical frequencies with the exact distribution.

Concretely, for each index $i = 1,\dots,100$, define the initial weights $w_i^{(0)} \;=\;\Bigl(2 + \tfrac{1}{10000}\,i\Bigr)^{1000}$. These weights are large enough to allow many decay steps. For each $t = 1,2,\dots,100$, we divide each weight once by its base factor, giving
$w_i^{(t)} \;=\; w_i^{(t-1)} \,\Big/\Bigl(2 + \tfrac{1}{10000}\,i\Bigr)$ and hence $ w_i^{(t)} = \Bigl(2 + \tfrac{1}{10000}\,i\Bigr)^{1000 - t}$. As $t$ increases, the distribution gradually flattens.

We stop at $t=100$ because methods that require infinite-precision real arithmetic already become unstable well before then. At each $t\in\{1,\dots,100\}$, we performed $10^6$ independent draws from the dynamic discrete distribution $\{\,w_i^{(t)}\}_{i=1}^{100}$ to estimate the empirical distribution.

To quantify agreement between empirical and theoretical distributions, we measured the Jensen--Shannon divergence \cite{lin1991} at each iteration. An exact sampler is not expected to achieve zero divergence because of Monte Carlo noise, but it should remain near that noise floor. Figure~\ref{fig:numerical-divergence} shows that EBUS stays at that baseline, whereas the inexact methods drift upward:

\begin{figure}[H]
\centering
\includegraphics[width=\linewidth]{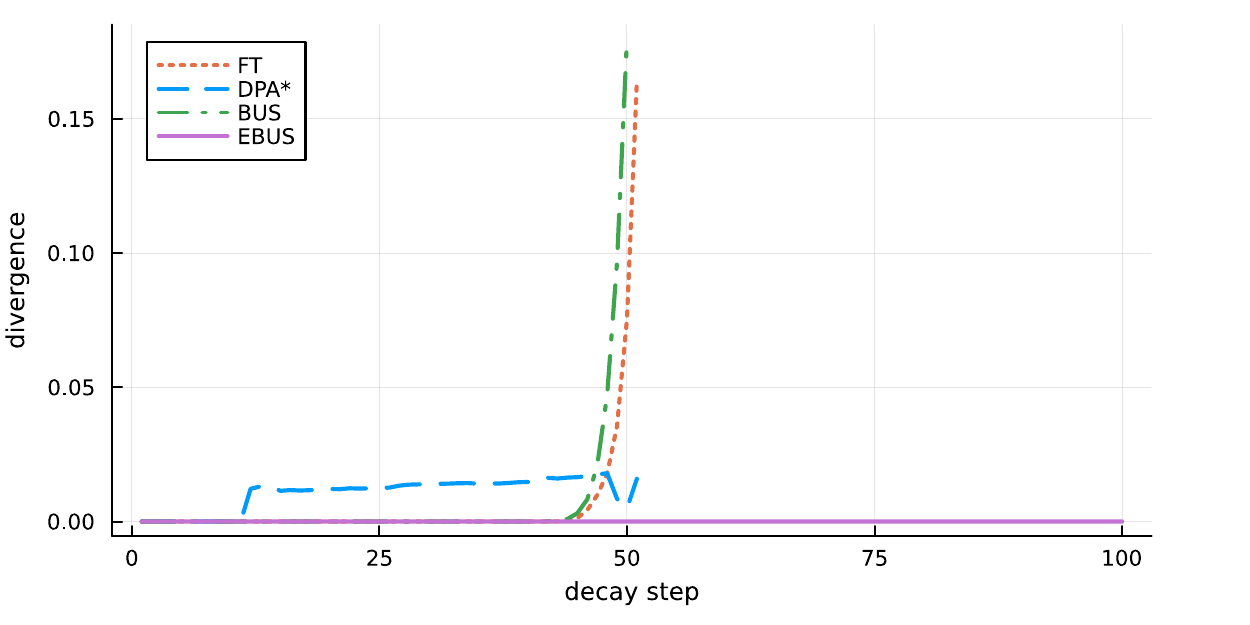}
\caption{Jensen-Shannon divergence between the empirical output distribution and the theoretical target distribution during the decay experiment. EBUS stays flat at 0 (up to sampling noise) whereas methods whose correctness requires infinite precision real arithmetic drift away from the exact distribution over time.}
\label{fig:numerical-divergence}
\end{figure}

\vspace{-5px}
Moreover, all inexact samplers failed beyond roughly 50 iterations, effectively entering a nonterminating or prohibitively slow state: within the allotted runtime, they could not return the next empirical frequency distribution. Figure~\ref{fig:numerical-step-50} shows that by decay step $t=50$ their empirical distributions already differ substantially from the true one, while EBUS still matches it up to ordinary sampling noise:

\begin{figure}[t]
\centering
\includegraphics[width=\linewidth]{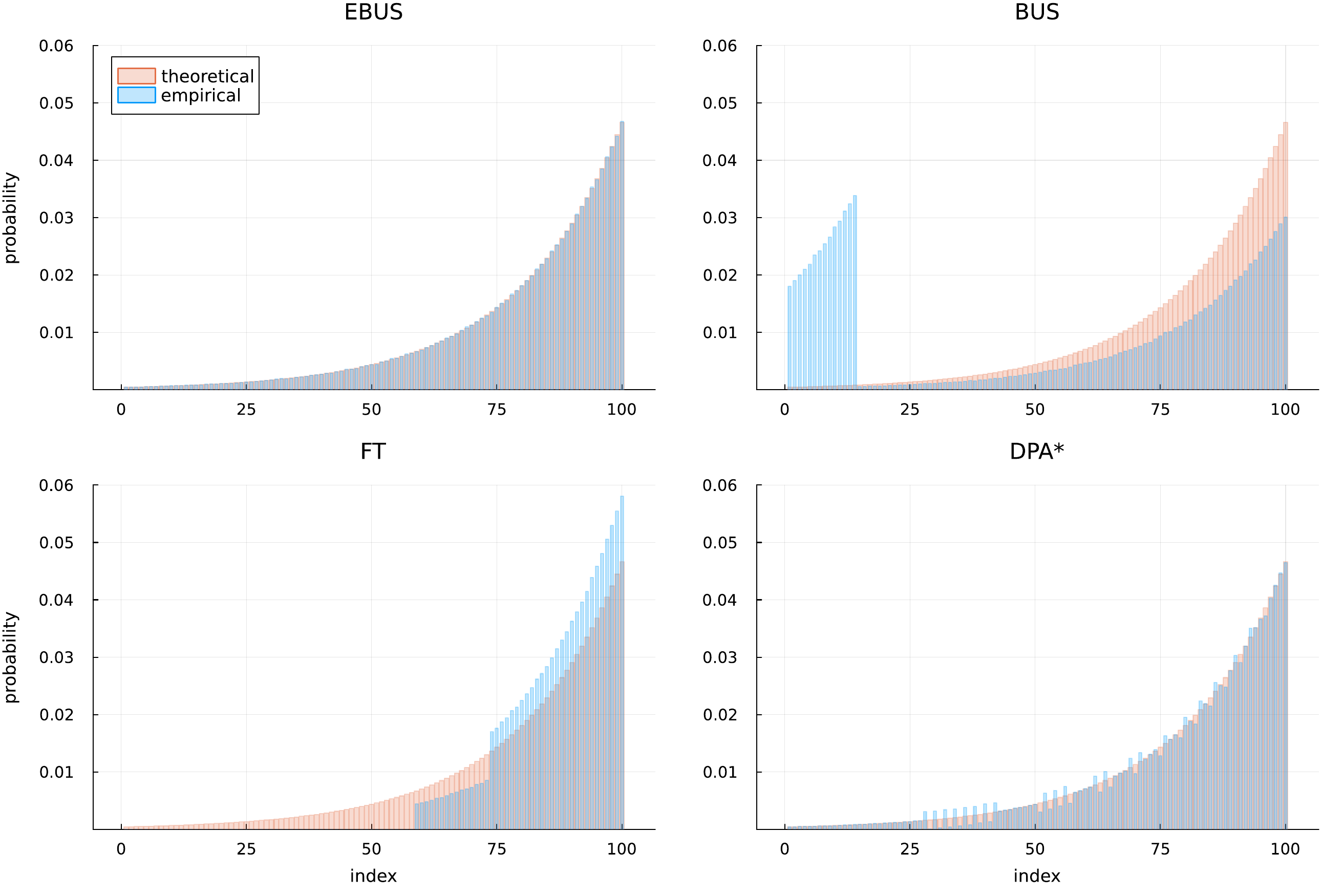}
\caption{Empirical vs. theoretical distributions at decay step $t=50$. EBUS remains aligned with it up to sampling noise, while the inexact samplers exhibit visible distortion.}
\label{fig:numerical-step-50}
\end{figure}

\subsection{Performance Comparison}

We evaluated performance across four scenarios, representing progressively higher levels of dynamicity:

\begin{itemize}
    \item \textsl{Static Sampling}: elements are added at initialization time, and then the distribution is fixed throughout the sampling process.
    \item \textsl{Dynamic Sampling with Fixed Range}: a weight of a random index is updated after one random draw from the distribution.
    \item \textsl{Dynamic Sampling with Decreasing Range}: a weight is removed after one random draw from the distribution until the sampler is resized to 1/10 of its initial size.
    \item \textsl{Dynamic Sampling with Increasing Range}: a weight is added after one random draw from the distribution until the sampler is resized to 10 times its initial size.
\end{itemize}

\begin{figure}[t]
\centering
\includegraphics[width=\linewidth]{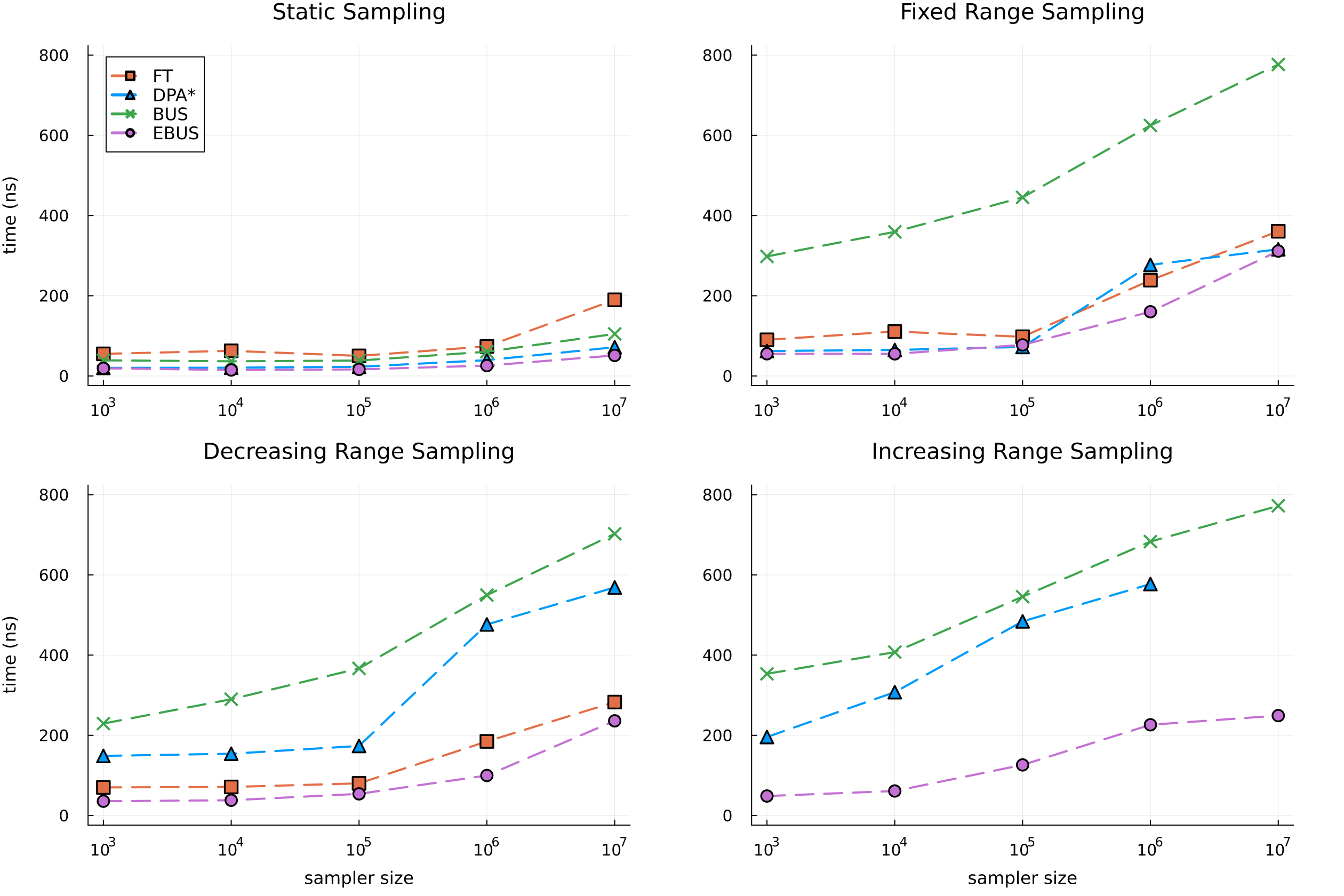}
\caption{Performance Benchmarks. Top Left (Static): Average time per sample vs. size. Top Right (Dynamic Fixed): Average time per iteration (1 sample + 1 update) on a fixed set of indices. Bottom Left (Dynamic Decreasing): Average time per iteration while the index set shrinks to 1/10th its initial size. Bottom Right (Dynamic Increasing): Average time per iteration while the index set grows to 10x its initial size. Plotted times are medians of 50 runs for sizes < \(10^6\), and 5 runs for larger sizes.}
\label{fig:performance-benchmarks}
\end{figure}

Throughout all experiments, each weight is sampled from the positive half of a standard Gaussian distribution. Figure~\ref{fig:performance-benchmarks} shows that our Julia implementation of EBUS is substantially faster than the original C++ BUS implementation and competitive with, or better than, the strongest inexact implementation we tested in all four scenarios. The more dynamic the benchmark, the larger EBUS's advantage. The upward trend in all EBUS benchmarks is consistent with some operations being memory-bound and therefore becoming more expensive as the sampler size grows. Additional benchmarks of uniform random sampling on vectors of increasing size, available in the benchmark repository, show a similar trend. The FT method does not support the dynamic increasing-range setting and therefore does not appear in that experiment. The DPA* method exceeded the available 16~GB of RAM when running the increasing-range benchmark with starting size \(10^7\), so it is omitted from the figure.

\section{Conclusion}

We presented EBUS (Exact BUcket Sampling), the first algorithm that combines exactness and strong performance for sampling from weighted dynamic distributions in finite precision. Achieving this required carefully designed sampling and update procedures that preserve exactness without sacrificing efficiency.

Our numerical experiments show that methods whose correctness requires infinite-precision real arithmetic can suffer large finite-precision errors in practice, enough to bias the sampling process or even halt it by breaking internal state. They also show that our implementation is competitive with the fastest previous implementations we found.

EBUS is therefore useful when finite-precision correctness matters, when efficiency matters, or both. Additionally, it can serve as a benchmark for exact methods on specific subclasses of distributions.


\bibliography{lipics-v2021-sample-article}
\appendix

\section{Deferred Proofs}\label{sec:proofs}

\boundaryrefinement*
\begin{proof}
Let \(M_\ell=W_\ell 2^G=SS_\ell 2^{\ell+G}\). Since \(SS_\ell\) is an integer, there is a smallest
\(D\ge 1\) such that \(M_\ell B^D\) is an integer. Therefore
\[
f_\ell=\{M_\ell\}=\sum_{m=1}^{D} d_m B^{-m}
\quad\text{with}\quad
d_m=\lfloor M_\ell B^m\rfloor \bmod B.
\]
These are exactly the digits computed by the algorithm, since at the \(m\)th iteration it uses
\[
t=\left\lfloor SS_\ell 2^{\ell+G+mb}\right\rfloor \bmod 2^b
=\lfloor M_\ell B^m\rfloor \bmod B
=d_m.
\]
At iteration \(m\), the loop samples an independent uniform digit
\(R_m\in\{0,\dots,B-1\}\) and compares it with \(d_m\). Thus the loop accepts at the first position where \(R_m<d_m\), rejects at the first position where \(R_m>d_m\), and also rejects if all \(D\) digits agree, because at that point the stopping condition \(s\ge 0\) is triggered. Define
\[
U=\sum_{m=1}^{D} R_m B^{-m}
=\frac{R_1B^{D-1}+R_2B^{D-2}+\cdots+R_D}{B^D}.
\]
Since \(R_1,\dots,R_D\) are independent and uniform on \(\{0,\dots,B-1\}\), every
\(D\)-tuple of digits occurs with probability \(B^{-D}\). Moreover, the map
\(
(r_1,\dots,r_D)\mapsto \frac{r_1B^{D-1}+\cdots+r_D}{B^D}
\)
is a bijection from \(\{0,\dots,B-1\}^D\) onto
\(
\left\{0,\frac1{B^D},\frac2{B^D},\dots,\frac{B^D-1}{B^D}\right\}.
\)
Hence \(U\) is uniform on that set. Recall that the loop accepts exactly when \(U<f_\ell\). Writing
\(
f_\ell=\frac{h}{B^D}
\)
for some
\(
h\in\{0,\dots,B^D-1\},
\)
exactly \(h\) of the \(B^D\) possible values of \(U\) are strictly smaller than \(f_\ell\). Hence
\(
\Pr[\text{accept}]=\Pr[U<f_\ell]=\frac{h}{B^D}=f_\ell,
\)
and therefore \(\Pr[\text{restart}]=1-f_\ell\).
\end{proof}

\intersamplingprob*
\begin{proof}
Write
\[
W_\ell 2^G = q_\ell + f_\ell,
\quad
q_\ell = \lfloor W_\ell 2^G \rfloor,
\quad
f_\ell \in [0,1).
\]
For every nonempty level we therefore have $A_\ell = q_\ell + 1$. In one outer iteration, the sampled integer lands in level $\ell$ with probability $A_\ell/A$. Conditioned on landing in level $\ell$, the sampler accepts immediately on the first $q_\ell$ points of that bucket and accepts on the final boundary point with probability exactly $f_\ell$ by Lemma~\ref{lem:boundary-refinement}. Hence the conditional acceptance probability for level $\ell$ is
\[
\frac{q_\ell + f_\ell}{q_\ell+1} = \frac{W_\ell 2^G}{A_\ell}.
\]
Multiplying by the probability \(A_\ell/A\) of landing in that bucket yields
\[
\frac{A_\ell}{A}\cdot\frac{W_\ell 2^G}{A_\ell} = \frac{W_\ell 2^G}{A}.
\]
Summing over levels gives the expression for $P_{\mathrm{acc}}$.
\end{proof}

\slowpathoverhead*
\begin{proof}
Let \(P_{\mathrm{boundary}}\) be the probability that one outer iteration enters the boundary-refinement path, and let \(P_{\mathrm{acc}}\) be the probability that
one outer iteration accepts some level. There is at most one boundary point per level, so
\(
P_{\mathrm{boundary}}\le \frac{N}{A}.
\)
By Lemma~\ref{lem:inter-level-one-round},
\(
P_{\mathrm{acc}}=\frac{M}{A}.
\)

Let \(S\) be the total number of boundary-refinement entries before termination.
The outer iterations are independent and repeat until the first accepting one.
Thus
\[
\mathbb{E}[S]
=
\sum_{i\ge 1}(1-P_{\mathrm{acc}})^{i-1}P_{\mathrm{boundary}}
=
\frac{P_{\mathrm{boundary}}}{P_{\mathrm{acc}}}
\le
\frac{N/A}{M/A}
=
\frac{N}{M}.
\]
\end{proof}

\intraproposals*
\begin{proof}
Each stored significand satisfies $\sigma_k \in [2^{b-1},2^b)$, and $B=2^{\lceil\log_2 n\rceil} < 2n$. Therefore the probability that one loop iteration accepts some index is
\[
\frac{1}{B2^b}\sum_{k=1}^n \sigma_k \ge \frac{1}{B2^b} \cdot n2^{b-1} > \frac{1}{4}.
\]
The number of iterations is geometric with success probability at least $1/4$, so its expectation is at most $4$.
\end{proof}

\coarseestimator*
\begin{proof}
Write
\(SS_{\lambda-r}=2^bH_r+L_r \) and \(0\le L_r<2^b\).
Then
\[
Y_\lambda
=\sum_{r\ge 0} SS_{\lambda-r}2^{-r+1-b}
=
\sum_{r=0}^{b} H_r2^{-r+1}
+
\sum_{r\ge b+1} H_r2^{-r+1}
+
\sum_{r\ge 0} L_r2^{-r+1-b}.
\]
The first sum is at least \(E\), since each floor rounds downward. The total loss from these \(b+1\) floors is less than \(b+1\). Also \(H_r<2^b\) for every \(r\), so
\[
\sum_{r\ge b+1} H_r2^{-r+1}
<
2^b\sum_{r\ge b+1}2^{-r+1}
=2,
\]
and similarly
\[
\sum_{r\ge 0} L_r2^{-r+1-b}
<
2^b\sum_{r\ge 0}2^{-r+1-b}
=4.
\]
Hence
\(
E \le Y_\lambda < E+(b+1)+2+4 = E+b+7
\). Finally, if the sampler is nonempty, then the top level \(\lambda\) contains at least one normalized significand, so \(SS_\lambda\ge 2^{b-1}\). Therefore
\(
Y_\lambda \ge 2^{1-b}SS_\lambda \ge 1.
\)
\end{proof}

\downwardbounds*
\begin{proof}
For every \(r\in S\),
\[
A_r(G') - 1
=
\left\lfloor W_r2^{G'}\right\rfloor
=
\left\lfloor \frac{W_r2^G}{2^d}\right\rfloor
=
\left\lfloor \frac{\lfloor W_r2^G\rfloor}{2^d}\right\rfloor
=
\left\lfloor \frac{A_r(G)-1}{2^d}\right\rfloor.
\]
Hence
\(
A_r(G')= \left\lfloor \frac{A_r(G)-1}{2^d}\right\rfloor+1.
\)
Summing over \(r\in S\) gives
\[
\sum_{r\in S} A_r(G')
=
|S|+\sum_{r\in S}\left\lfloor \frac{A_r(G)-1}{2^d}\right\rfloor
\le
|S|+\left\lfloor \frac{\sum_{r\in S}(A_r(G)-1)}{2^d}\right\rfloor,
\]
If \(d\ge 1\), then
\[
\sum_{r\in S} A_r(G')
\le
|S|+\left\lfloor \frac{\sum_{r\in S}A_r(G)-|S|}{2}\right\rfloor.
\]
Finally, if \(S\) is the set of all nonempty levels, then
\[
A(G')
=
\sum_{r\in S} A_r(G')
\le
|S|+\left\lfloor \frac{A(G)-|S|}{2^d}\right\rfloor
\le
N+\left\lfloor \frac{A(G)}{2^d}\right\rfloor,
\]
\end{proof}

\scanlength*
\begin{proof}
Let \(A(G)=\sum_\ell A_\ell(G)\). Before an outer iteration starts, the pre-sampling check of Theorem~\ref{thm:global-shift-adjustments} ensures that \(A(G)\ge 2^K\): if initially \(A(G)<2^K\), then the lazy increase is performed first. Also, for any safe shift, the argument in the proof of Theorem~\ref{thm:max-updated-levels} together with Lemma~\ref{lem:unit-bucket-threshold} shows that all but at most the top \(2b\) levels have summary \(0\) or \(1\). Thus, after those first \(2b\) levels, any remaining nonempty level is a unit bucket, chosen with probability at most \(1/A(G)\le 2^{-K}\). Let \(R\) be the number of levels below those top \(2b\) levels, so \(R\le N\). If the nonempty unit buckets in that suffix occur at positions
\(
1\le p_1<\cdots<p_r\le R,
\)
then the expected number of additionally scanned levels is
\[
\frac{p_1+\cdots+p_r}{A(G)}
\le
\frac{1+2+\cdots+R}{2^K}
\le
\frac{1+2+\cdots+N}{2^K}
=
\frac{N(N+1)}{2^{K+1}}.
\]
Adding the first \(2b\) levels gives the claim.
\end{proof}

\section{Efficient Restoration of Level Summaries}\label{sec:restoration-levels}
The shift policy of Theorem~\ref{thm:global-shift-adjustments} specifies only the new value of \(G\). In the main text we showed that any safe shift change can affect at most \(2b\) level summaries. In this appendix, we explain how to restore the exact bucket sizes \(A_\ell(G)\) on precisely those candidate levels more efficiently than recomputing each of them from the full exact sums.

Recall from the main text that levels below the threshold
\(
\theta(G,z) = -b-G-\lfloor \log_2(z) \rfloor,
\)
need no update. For every shift \(G\), define also the second threshold
\(
\eta(G) = -b-G.
\)
It will be proved that for an upward shift \(\eta(G)\) marks the first level for which the lower \(b\) bits of the exact significand sum may affect the bucket size.

We first record the two additional facts used by the repair routines.

\begin{lemma}[Exact rescaling under a downward shift]
\label{lem:exact-downward-shift}
Let \(G'\le G\). Then for every level \(\ell\),
\[
A_\ell(G')=
\begin{cases}
0, & SS_\ell=0,\\[1mm]
\left\lfloor \dfrac{A_\ell(G)-1}{2^{G-G'}}\right\rfloor+1, & SS_\ell>0.
\end{cases}
\]
\end{lemma}

\begin{proof}
If \(SS_\ell=0\), then \(W_\ell=0\) and both sides are zero.
Assume \(SS_\ell>0\), and write
\[
W_\ell 2^G = q+f,
\quad
q=\lfloor W_\ell 2^G\rfloor,
\quad
f\in[0,1).
\]
Then \(A_\ell(G)=q+1\), so \(A_\ell(G)-1=q\). If \(\delta=G-G'\ge 0\), then
\[
A_\ell(G')-1
=
\left\lfloor W_\ell 2^{G'}\right\rfloor
=
\left\lfloor \frac{q+f}{2^\delta}\right\rfloor
=
\left\lfloor \frac{q}{2^\delta}\right\rfloor,
\]
because \(0\le f/2^\delta<1\). This is exactly the stated formula.
\end{proof}

\begin{lemma}[Upper-half recomputation]
\label{lem:upper-half-recomputation}
Write
\(
SS_\ell = 2^b H_\ell + L_\ell
\) for
\(
0\le L_\ell < 2^b
\).
If
\(
\ell < \eta(G) = -b-G,
\)
then
\[
A_\ell(G)=
\begin{cases}
0, & SS_\ell=0,\\[1mm]
\left\lfloor H_\ell 2^{\ell+G+b}\right\rfloor+1, & SS_\ell>0.
\end{cases}
\]
In particular, for such levels the lower \(b\) bits \(L_\ell\) do not affect \(A_\ell(G)\).
\end{lemma}

\begin{proof}
If \(SS_\ell=0\), there is nothing to prove. Otherwise,
\[
W_\ell 2^G
=
SS_\ell 2^{\ell+G}
=
(2^bH_\ell+L_\ell)2^{\ell+G}
=
H_\ell 2^{\ell+G+b}+L_\ell 2^{\ell+G}.
\]
Since \(\ell<-b-G\), we have \(\ell+G\le -b-1\), hence
\(
0\le L_\ell 2^{\ell+G}<1.
\)
Therefore
\(
\lfloor W_\ell 2^G\rfloor =
\left\lfloor H_\ell 2^{\ell+G+b}\right\rfloor,
\)
which yields the claimed formula.
\end{proof}

Lemma~\ref{lem:unit-bucket-threshold} from the main text and Lemma~\ref{lem:upper-half-recomputation} identify two nested low-level regions:
\(
\ell<\theta(G,z)
\Longrightarrow\
A_\ell(G)\in\{0,1\}
\)
and
\(
\theta(G,z)\le \ell<\eta(G)
\Longrightarrow
A_\ell(G)\)
depends only on the upper \(b\) bits of \(SS_\ell\). Only levels \(\ell\ge \eta(G)\) may require the full exact sums.

We now use the previous results to describe repairs after decreasing or increasing the shift. The next two corollaries show how to efficiently update level summaries after a shift.

\begin{corollary}[Repair after a downward shift]
\label{cor:repair-downward-shift}
Suppose the shift decreases from \(G\) to \(G'<G\). Then every level \(\ell\ \ge \theta(G,z)\) falls into exactly one of the following two cases.

\begin{enumerate}
    \item If
    \(
    \theta(G,z)\le \ell<\theta(G',z),
    \)
    then
    \[
    A_\ell(G')=
    \begin{cases}
    0, & SS_\ell=0,\\
    1, & SS_\ell>0.
    \end{cases}
    \]
    Hence these levels are exactly the ones that must be flattened to unit buckets.

    \item If
    \(
    \theta(G',z)\le \ell\le \lambda,
    \)
    then
    \[
    A_\ell(G')
    =
    \begin{cases}
    0, & SS_\ell=0,\\[1mm]
    \left\lfloor \dfrac{A_\ell(G)-1}{2^{G-G'}}\right\rfloor+1, & SS_\ell>0.
    \end{cases}
    \]
    Hence these levels are updated by Lemma~\ref{lem:exact-downward-shift}.
\end{enumerate}

\end{corollary}

\begin{proof}

If \(\theta(G,z)\le \ell<\theta(G',z)\), then Lemma~\ref{lem:unit-bucket-threshold} applies to the new shift \(G'\), yielding exactly the stated unit-bucket formula. Finally, if \(\theta(G',z)\le \ell\le\lambda\), then the new shift does not force the level into the unit-bucket regime, so Lemma~\ref{lem:exact-downward-shift} gives the stated formula.
\end{proof}

\begin{corollary}[Repair after an upward shift]
\label{cor:repair-upward-shift}
Suppose the shift increases from \(G\) to \(G'>G\). Then, every level $\ell \ge \theta(G', z) $ falls into exactly one of the following two cases.

\begin{enumerate}

    \item If
    \(
    \theta(G',z)\le \ell<\eta(G'),
    \)
    then
    \[
    A_\ell(G')=
    \begin{cases}
    0, & SS_\ell=0,\\[1mm]
    \left\lfloor H_\ell 2^{\ell+G'+b}\right\rfloor+1, & SS_\ell>0,
    \end{cases}
    \quad
    H_\ell=\left\lfloor \frac{SS_\ell}{2^b}\right\rfloor.
    \]
    Hence these levels can be recomputed using only the upper \(b\) bits of \(SS_\ell\).

    \item If
    \(
    \eta(G')\le \ell\le\lambda,
    \)
    then
    \[
    A_\ell(G')=
    \begin{cases}
    0, & SS_\ell=0,\\[1mm]
    \lfloor W_\ell 2^{G'}\rfloor+1, & SS_\ell>0.
    \end{cases}
    \]
    Hence these levels are recomputed from the full exact sums.
\end{enumerate}
\end{corollary}

\begin{proof}
If \(\theta(G',z)\le \ell<\eta(G')\), then \(\ell<\eta(G')\), so Lemma~\ref{lem:upper-half-recomputation} applies and shows that only the upper \(b\) bits of \(SS_\ell\) matter. The remaining levels satisfy \(\eta(G')\le \ell\le\lambda\). For them neither simplification is available, so one restores the exact value directly from the full exact sum. Levels above \(\lambda\) are empty and remain zero.
\end{proof}

\section{Bulk Sampling}
\label{cor:bulk-sampling}

The scalar sampler described in the main text produces one exact sample at a time. When many samples are needed from the same current distribution, it is convenient to separate the choice of the exponent level from the choice of the item inside that level.

Fix the current sampler state, and let $L_1,\dots,L_N$ denote the nonempty levels. Write
\(
W=\sum_{\ell=1}^N W_\ell
\)
and
\(
p_\ell = \frac{W_\ell}{W}.
\)
For an item $j\in L_\ell$, define its conditional within-level probability by
\(
q_{\ell,j}
=
\frac{\sigma_j}{\sum_{k\in L_\ell}\sigma_k}.
\)
By Theorem~\ref{thm:inter-level-exactness} and Theorem~\ref{thm:intra-level-exactness}, one scalar call first chooses a level $\ell$ with probability $p_\ell$ and then, conditioned on that level, chooses item $j\in L_\ell$ with probability $q_{\ell,j}$. We now describe how to generate multiple samples at once.

\begin{lemma}[Bulk decomposition]
\label{lem:bulk-decomposition}
Let $m\ge 1$. The joint distribution of $m$ independent scalar samples is
obtained as follows:
\begin{enumerate}
    \item draw a level-count vector
    \(
    (C_1,\dots,C_N)\sim \mathrm{Multinomial}(m; p_1,\dots,p_N)
    \).
    \item for each level $\ell$, draw $C_\ell$ independent samples from the
    conditional distribution $(q_{\ell,j})_{j\in L_\ell}$.
\end{enumerate}
\end{lemma}

\begin{proof}
Let $Z_1,\dots,Z_m$ be the level labels produced by $m$ independent scalar calls. Then $Z_1,\dots,Z_m$ are i.i.d.\ with
\(
\Pr[Z_t=\ell]=p_\ell,
\)
so their count vector is multinomial with parameters $(m;p_1,\dots,p_N)$. Conditioned on the realized level labels, the item chosen at each position depends only on its level and has conditional law $(q_{\ell,j})_{j\in L_\ell}$, independently across positions. Finally, conditioned on the counts $(C_1,\dots,C_N)$, the level-label sequence is uniform over all sequences with those counts.
\end{proof}

The remaining task is therefore to sample the multinomial count vector $(C_1,\dots,C_N)$ exactly.

\begin{lemma}[Sequential binomial construction]
\label{lem:sequential-binomial-multinomial}
Define $R_1=m$, and sample
\(
C_1 \sim \mathrm{Binomial}(R_1,p_1).
\)
Then, for $\ell=2,\dots,N-1$, let
\(
R_\ell = m-\sum_{r<\ell} C_r
\)
and sample
\[
C_\ell \sim \mathrm{Binomial}\!\left(
    R_\ell,\,
    \frac{p_\ell}{1-\sum_{r<\ell} p_r}
\right)
=
\mathrm{Binomial}\!\left(
    R_\ell,\,
    \frac{W_\ell}{\sum_{r\ge \ell} W_r}
\right).
\]
Finally set
\(
C_N = m-\sum_{r=1}^{N-1} C_r.
\)
Then $(C_1,\dots,C_N)$ has the multinomial distribution with parameters $(m;p_1,\dots,p_N)$.
\end{lemma}

\begin{proof}
This is the standard binomial decomposition of the multinomial distribution. At stage $\ell$, after the previous counts have been fixed, there remain exactly $R_\ell$ unresolved samples, and each of them belongs to level $\ell$ with conditional probability
\[
\frac{p_\ell}{1-\sum_{r<\ell}p_r}
=
\frac{W_\ell}{\sum_{r\ge \ell}W_r}.
\]
Hence $C_\ell$ has the stated binomial law, and multiplying the corresponding conditional probabilities over $\ell=1,\dots,N-1$ yields exactly the multinomial mass function.
\end{proof}

Lemma~\ref{lem:sequential-binomial-multinomial} reduces exact multinomial sampling to a sequence of exact binomial samplers. For this step, one may use the exact sublinear binomial sampler of Farach-Colton and Tsai \cite{colton2015}. Their algorithm assumes that the binomial parameter can be queried exactly, bit by bit, in constant time. In our setting this assumption is satisfied in the word RAM model, because each parameter is an exact rational of the form
\(
\frac{W_\ell}{\sum_{r\ge \ell} W_r}.
\)

Moreover, this bulk path does not require arbitrary-precision integers. For an IEEE $b$-bit binary format with $N$ possible normalized levels, after multiplying all level weights by a common power of two corresponding to the smallest level exponent, every level weight and every suffix sum $\sum_{r\ge \ell}W_r$ becomes an exact nonnegative integer. Such an integer consists of:

\begin{itemize}
    \item up to $2b$ bits for the exact significand sum.
    \item at most $N$ bits for the exponent offset, since there are $N$
    possible normalized levels in the chosen IEEE format.
\end{itemize}

Therefore fewer than $2b+N$ bits are sufficient, so a fixed-width integer type of that size is enough for all exact integer quantities used in the bulk multinomial stage. In particular, this is a fixed-width computation, not a general big-integer one. For the IEEE-double implementation, this yields fewer than $128+2098=2226$ bits, so a fixed $2240$-bit integer type suffices; with a 64-bit word, this is $35$ words.

Combining Lemma~\ref{lem:bulk-decomposition} with
Lemma~\ref{lem:sequential-binomial-multinomial} yields an exact bulk-sampling routine: first sample the exact multinomial level counts using exact binomial draws, then sample within each level according to the exact conditional law. The resulting batch has exactly the same distribution as $m$ repeated independent scalar calls, while batching the inter-level work across the whole sample.

\end{document}